%% file: burke.tex
\documentclass[twocolumn,preprintnumbers,aps,eqlabels,prl]{revtex4}

\usepackage{graphicx}
\usepackage{dcolumn}
\usepackage{bm}

\input dft

\input epsf
\input psfig
\input rotate

\begin{document} 

\title{
Time-dependent density functional theory:
Past, present, and future}
\author{Jan Werschnik and E.K.U. Gross}
\affiliation{Institut f\"ur Theoretische Physik, Freie Universit\"at Berlin, 
Arnimallee 14, 14195 Berlin, Germany}
\author{Kieron Burke}
\affiliation{Department of Chemistry and Chemical Biology, Rutgers University, 
610 Taylor Road, Piscataway, NJ 08854}
\date{\today}

\begin{abstract}
Time-dependent density functional theory (TDDFT) is presently enjoying 
enormous popularity in quantum chemistry, as a useful tool for 
extracting electronic excited state energies.
This article discusses how TDDFT is much broader in scope, and
yields predictions for many more properties.
We discuss some of the challenges involved in making accurate
predictions for these properties.
\end{abstract}
\maketitle

Kohn-Sham density functional theory
\cite{HK64,KS65,DG90} is the method of choice
to calculate ground-state properties of large molecules, because it replaces
the interacting many-electron problem with an effective
single-particle problem that can be solved much faster.
Time-dependent density functional theory (TDDFT) applies
the same philosophy to time-dependent problems.
We replace the complicated many-body time-dependent Schr\"odinger
equation by a set of time-dependent single-particle equations whose
orbitals yield the same time-dependent density $n(\br,t)$. 
We can do this because the Runge-Gross theorem\cite{RG84}
proves that, for a given initial wavefunction,
particle statistics and interaction, a given time-dependent
density $\n(\br, t)$ can arise from at most one time-dependent external
potential $v\ext(\br, t)$.  We define time-dependent Kohn-Sham (TDKS) equations
that describe $N$ non-interacting electrons that evolve in $v\s(\br, t)$,
but produce the same $\n(\br, t)$ as that of the interacting
system of interest.
Development and applications of TDDFT have enjoyed exponential growth in 
the last few years\cite{MG04,FB04,MBAG01,GDP96},
and we hope this merry trend will continue. 

The scheme yields predictions for a huge variety of phenomena, 
that can largely be classified into three groups:
(i) the non-perturbative regime, with systems in laser fields
so intense that perturbation theory fails,
(ii) the linear (and higher-order) regime, which yields the
usual optical response and electronic transitions, and
(iii) back to the ground-state, where the fluctuation-dissipation
theorem produces {\em ground-state} approximations from TDDFT
treatments of excitations.

\begin{figure}[htb]
\unitlength1cm
\includegraphics*[scale=1.00, angle=0]{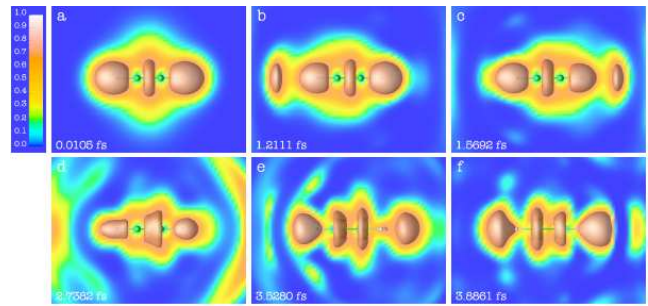}
\caption{Snapshots of the time-dependent ELF for the excitation of acetylene by a 17.5 eV laser pulse\cite{WG04}. The pulse had a total length of 7 fs, an intensity of $1.2 \times 10^{14} \mbox{W} \mbox{cm}^{-2}$, and was polarized along the molecular axis. Ionization and the transition from the bonding $\pi$ state to the anti-bonding $\pi^*$ state are clearly visible.}
\label{fig:tdelf}
\end{figure}

In the first, {\bf non-perturbative regime},
we have systems in intense laser fields with electric field
strengths that are comparable to or even exceed the attractive 
Coulomb field of the nuclei\cite{MG04}.   The time-dependent field cannot
be treated perturbatively, and even solving the time-dependent
Schr\"odinger equation for the evolution of
two interacting electrons is barely feasible with 
present-day computer technology\cite{PMDT00}.
For more electrons in a time-dependent field,
wavefunction methods are 
prohibitive, and in the regime of (not too high) laser intensities, 
where the electron-electron interaction is
still of importance TDDFT is essentially the only practical scheme
available.
With the recent advent of
atto-second laser pulses, the electronic time-scale has become accessible. 
Theoretical tools to analyze the dynamics of excitation processes on
the attosecond time scale will become more and more important.
An example of such a tool is the time-dependent electron
localization function (TDELF) \cite{BMG04,EGV04}. This quantity
allows the {\bf time-resolved} observation of the {\bf formation, 
modulation, and breaking of chemical bonds},
thus providing a visual understanding of the dynamics of excited electrons 
(for an example see Fig.\ref{fig:tdelf} 
and Ref. \cite{WG04}).
The natural way of calculating the TDELF is from the TDKS orbitals. 
Recent applications in the 
beyond perturbative regime range from above-threshold ionization
of metal clusters\cite{NBU04} to coherent control of 
quantum wells\cite{WU04}
to multiharmonic generation in benzene\cite{BNZM03}.

A much larger group of applications in chemistry is
the linear response to
a spatially uniform electric field, i.e., the {\bf optical response}
in the dipole approximation\cite{PGG96,C96}.  Formal analysis of this situation
shows that TDDFT yields predictions for electronic excitations, both
their position (transition frequency) and intensity (oscillator
strength).  These are corrections to transitions
between occupied and unoccupied levels of the ground-state KS potential,
thus providing a simple interpretation of those levels\cite{AGB03}. 
In the area of calculating electronic excitations, TDDFT is rapidly becoming 
a standard tool, complimentary to existing wavefunction techniques\cite{FA02}.
Just as in the ground-state case, it has the advantage in computational speed,
allowing study of larger systems than with traditional methods, and the
usual disadvantage (or excitement) of being unsystematic and artful. A final
application is to write the {\bf  ground-state XC energy} in terms of 
the frequency-dependent response function, and so linear
response TDDFT yields
approximate treatments of
the ground-state problem\cite{LGP00,ALL96,FNGB05,DRSL04}.

A random walk through some of 2004's papers using TDDFT gives
some feeling for the breadth of applications.  Most are in the
linear response regime.  In inorganic chemistry, the optical response
of many transition metal
complexes\cite{PCLK05,VSHL04,CHBG04,SVR04,AFS04,JAZ03} has been calculated,
and even some X-ray absorption\cite{FSRD04}.
In organic chemistry, the response of thiouracil\cite{SL04}
and s-tetrazine\cite{OKH04},
and annulated porphyrins\cite{RNHM03} were investigated. 
In photobiology, potential energy curves for the trans-cis
photo-isomerization of protonated Schiff base of retinal\cite{TI04}
have been calculated.
For these and other systems, there is great interest in
charge-transfer excitations\cite{BSH04,JC04,DH04,RF04,T03,JL03}, but (as we
later discuss) intermolecular charge transfer is a demanding problem for TDDFT.
Another major area of application is clusters, large and small, covalent and
metallic, and everything 
inbetween\cite{DHSM04,GGJI04,BS04,PRS04,RSCG04,DRGM04,BNW04},
including Met-Cars\cite{MCRP04}.
Several studies include solvation, for example, the behavior of metal ions
in explicit water\cite{BBSV04}.
TDDFT in linear response
can also be used to calculate both electronic and magnetic
circular dichroism\cite{SMBS04,SZBA04},
and has been applied to helical aromatics\cite{WSTI04},
and to artemisinin complexes in solution\cite{MMME04}.  
There have also been applications in materials\cite{WAT04,GKPE04} and quantum 
dots\cite{HMW04}
but, as discussed
below, the
optical response of solids requires some non-local approximations\cite{BSVO04}.
Beyond the linear regime, there is also
growing interest in second- and third-order
response\cite{IASM04,JZ04,KMT04,MT04} in all these fields.

A wonderful aspect of TDDFT is that a single approximation to
the time-dependent XC potential implies predictions for all these
quantities.  This is analogous to the ground-state case, where
a single approximation to $E\xc$ can be applied to all ground-state properties
of all electronic systems, such as dissociation energies, bond lengths
and angles, vibrational frequencies, etc., of atoms, molecules, clusters,
and solids.  The starting point of most TDDFT approximations is the
adiabatic local density approximation (ALDA), which approximates
the XC potential at point $\br$ and time $t$
by that of a ground-state uniform electron gas
of density $n(\br t)$.   This is clearly accurate when the density varies
sufficiently slowly in time and space, but works surprisingly well
beyond that limit for many systems and properties,
just as LDA does for most ground-state problems.

We make an important distinction here between the mature subject of
ground-state DFT, and the developing one of TDDFT.  In the former,
there is general consensus on which properties are captured by which
functionals, and the aim is
toward higher accuracy\cite{TPSS03}.  One expects chemical
bonds to form in modern KS DFT calculations, and one hopes to use better
functionals to produce better accuracy and reliability\cite{PRTS04}.  
But time-dependent quantum mechanics probes a far more diverse
range of electronic phenomena, and 
in TDDFT, we
are still exploring even which properties
are captured at all by the presently available approximate functionals.  Quantitative
accuracy is less of an issue as yet.
Most data on the performance of TDDFT are for systems driven by some
external field. Practically nothing is known about how TDDFT performs in
the description of relaxation processes, i.e.,
on the time evolution of large systems starting
from a non-equilibrium initial condition\cite{BCG05}. 
A closely related question is the description
of density fluctuations within TDDFT\cite{Mb04}.

At this point, we introduce a few equations, to make
the discussion more precise.
We use
atomic units throughout, and suppress spin indices.
For brevity, we
drop commas between arguments wherever the meaning is clear.
 We write the TDKS
equations as
\ben
i \frac{d\phi_j (\br t)}{dt}  = 
\left( - \frac{\nabla^{2}}{2} + v\s[n](\bx) \right)
\phi_{j}(\bx)\quad.
\label{TDKS}
\een
whose density 
$n(\bx) = \sum_{j=1}^{N} | \phi_{j}(\bx)|^{2}$
is precisely that of the real system.   We {\em define} the
exchange-correlation potential via
\ben
v\s(\bx) = v\ext(\bx) + 
\int d^3r'\; \frac{\n(\br't)}{|\br-\br'|} + v\xc (\bx).
\label{vxc}
\een
The exchange-correlation potential, $v\xc(\bx)$ 
is in general a functional of the entire history
of the density, $\n(\bx)$, the initial interacting wavefunction
$\Psi(0)$, and the initial Kohn-Sham wavefunction, $\Phi(0)$.
This functional is a very complex one, much more so than the
ground-state case.   Knowledge of it implies solution of all
time-dependent Coulomb interacting problems.
If we always begin in a non-degenerate
ground state\cite{MBW02}, the initial-state dependence can be subsumed by
the Hohenberg-Kohn theorem\cite{HK64}, and the only unknown part
of $v\s(\br t)$, the exchange-correlation (XC) potential, is a functional of
$n(\br t)$ alone.

In the special case of the response of the ground
state to a weak external field, the system's
response is characterized by the non-local susceptibility
\ben
\label{chi}
\delta \n(\bx) = \int dt' \int d^3r' 
\chi[\n_0](\br,\br';t-t')\ \delta v\ext(\br' t').
\een
$\chi$ is a functional of the {\em ground-state} density, $\n_0(\br)$.
The central equation of TDDFT linear response\cite{PGG96} is a Dyson-like
equation for the true $\chi$ of the system:
\bea
\chi(\br\br'\omega)
&=& \chi\s(\br\br'\omega)
+ \int d^3r_1\int d^3r_2\ \chi\s(\br\br_1\omega)\nonumber\\
&\times&
\left(\frac{1}{|\br_1-\br_2|}+f\xc (\br_1\br_2\omega) \right)
\chi(\br_2\br'\omega),
\label{Dyson}
\eea
Here $\chi\s$ is the {\em Kohn-Sham} response function, constructed
from KS energies and orbitals:
\ben
\chi\s(\br\br'\omega) = 2 \sum_{q}
\frac{\Phi_q(\br)\ \Phi_q^*(\br')}
{\omega-\omega_q+i 0_+} + c.c.(\omega\to-\omega)
\label{chis}
\een
where $q$ is a double index, representing a transition from
occupied KS orbital $i$ to unoccupied KS orbital $a$,
$\omega_q=\epsilon_a-\epsilon_i$, 
and $\Phi_q(\br) = \phi_i^*(\br)\phi_a(\br)$.  Thus $\chi\s$ is
purely a product of the ground-state KS calculation.  On the other
hand, the XC kernel is defined as
\ben
f\xc[\n_0](\br\br',t-t') = \delta v\xc(\br t)/\delta \n(\br' t')|_{\n_0}.
\label{fxc}
\een
This is a much simpler quantity than $v\xc[\n](\bx)$, since
the functional is only evaluated at the ground-state density.  It is
non-local in both space and time.  The non-locality in time
manifests itself as a frequency dependence in the Fourier transform,
$f\xc(\br\br'\omega)$.

Next, we introduce Casida's equations\cite{C96}, in which the poles of
$\chi$ are found as the solution to an eigenvalue problem:
\begin{equation}
\sum_{q'} {\tilde\Omega}_{qq'} (\omega)\ a_{q'} = \omega^2\ a_q,
\label{Casida}
\end{equation}
where
\ben
{\tilde\Omega}_{qq'}(\omega)
= \omega^2_q \delta_{qq'} + 2 {\sqrt{\omega_q\omega'_q}}
\langle q | f\Hxc (\omega) | q' \rangle.
\label{Odef}
\een
and $
\langle q | f\Hxc (\omega) | q' \rangle$ is the matrix element
of the (Hartree)-XC kernel in the set of functions $\Phi_q(\br)$.
Eigenvalues yield the square of transition frequencies, while 
eigenvectors yield oscillator strengths. Ignoring off-diagonal matrix elements
can yield much insight into the nature of the TDDFT corrections to
the KS transitions\cite{AGB03}.

Lastly, we mention how TDDFT produces sophisticated approximations
to the {\em ground-state} exchange-correlation energy.  The adiabatic
connection fluctuation-dissipation formula is:
\bea
E\xc[\n_0]&=&-\half \int_0^1\; d\lambda\; \int d^3r\; \int d^3r'\;
\frac{1}{|\br-\br'|}
\nonumber\\
&&\!\!\!\!\!\! \int_0^\infty\frac{d\omega}{\pi}
\left\{\chi\l[\n_0](\br\br'\omega) + n_0(\br) \delta^{(3)}(\br-\br')\right\}
\label{Exc}
\eea
where the coupling-constant $\lambda$ is defined to multiply the
electron-electron repulsion in the Hamiltonian, but the external
potential is adjusted to keep the density fixed\cite{LP75,GL76}.
So any model for $f\xc$, even setting it to zero (called the
Random Phase Approximation), yields a sophisticated approximation to
$E\xc$, by solving Eq. (\ref{Dyson}) for $\chi$ (at each $\lambda$)
and inserting in Eq. (\ref{Exc}).

All the above equations are formally exact.  In any practical DFT
calculation, approximations must be made.
The most common approximation in TDDFT is the {\em adiabatic}
approximation, in which
\ben
v\xc\adia [\n](\br t) = v\xc^{\rm gs} [\n_0] (\br)|_{\n_0(\br)=\n(\br t)},
\label{adia}
\een
i.e., the XC potential at any time depends only on the density
at that time, not on its entire history.  This becomes
exact for slow variations in time. Most applications, however, are not in this 
slowly varying regime. Nevertheless, results obtained within the adiabatic
approximation are, in most cases, rather accurate. Any ground-state
approximation (LDA, GGA, hybrid) automatically provides an adiabatic
approximation (e.g., ALDA) in TDDFT.  Moreover, the XC kernel
is frequency-independent
in the adiabatic approximation, taking its $\omega\to 0$ value.

As mentioned above, TDDFT is proving very useful in predicting
optical response properties of molecules.
The Casida equations have been encoded in most standard quantum
chemical packages, and efficient algorithms developed to 
extract the lowest-lying excitations.   A small survey is given
by Furche and Ahlrichs\cite{FA02}.  Typical chemical
calculations are done with the B3LYP\cite{B93} functional, and
typical results are transition
frequencies within 0.4 eV of experiment, and structural properties
of excited states 
are almost as good as those of ground-state calculations
(bond lengths to within 1\%,
dipole moments to within 5\%, vibrational frequencies to within 5\%).
Most importantly, this level of accuracy appears sufficient in
most cases to qualitatively identify the nature of the most intense
transitions, often debunking cruder models that have been used for
interpretation for decades.  This is proving especially useful for
the photochemistry of biological molecules\cite{MLVC03}.
An alternative implementation, often favored by physicists,
is to propagate the TDKS equations in real time, having given
the system an initial weak perturbation. Such calculations
either use a real-space-grid\cite{VOC99,MCBR03} 
or plane waves\cite{CPMD95}.

This article is {\em not} about the (admittedly) gratifying successes of 
TDDFT calculations, which are dicussed in recent reviews\cite{MG04,FB04} and the
recent literature.  We begin from there, and explore a much wider
arena.
\begin{figure}[htb]
\unitlength1cm
\begin{picture}(6,6)
\put(-2.5,13.5){
\includegraphics{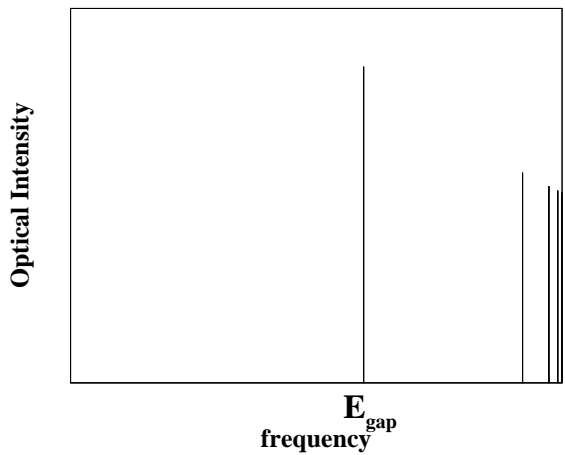}}
\end{picture}
\caption{Cartoon of the exact optical absorption spectrum of an atom or molecule,
with discrete transitions represented by straight lines (see text).}
\label{f:exact}
\end{figure}
To do this, in Fig. \ref{f:exact} we have drawn a cartoon
(literally, a stick figure) to 
represent the information in a typical calculation.
Each line represents a transition, with its position denoting the transition
frequency and its height
proportional to the oscillator strength.  For the He atom, the 
2s$\to$2p singlet transition is at 21.2 eV, while for the
N$_2$ molecule, the $^3\Pi\g$ transition is at 7.4 eV. 
Many applications of TDDFT report only the positions of the few lowest
optically-allowed transitions, while some report also their oscillator
strengths.

To help our analysis, we
list qualitatively different sources of error in the
predictions from any TDDFT calculation.  We refer to
them as the four deadly sins:
\begin{itemize}

\item The sin of the {\bf ground-state}:
Errors in the underlying {\em ground-state} DFT calculation.
If the KS orbital energies are wrong to begin with, TDDFT corrections
cannot produce accurate results.

\item
The sin of {\bf locality}:
Errors due to local (or gradient-corrected) approximations to an adiabatic
$f\xc(\br\br')$, i.e., properties that require non-locality in $|\br-\br'|$.

\item
The sin of {\bf forgetfulness}:
Phenomena missing when the adiabatic approximation is made, i.e.,
properties that require non-locality in time, i.e., memory.

\item 
The sin of the {\bf wavefunction}:
Even if the {\em exact} $v\xc(\bx)$ is used, solution of the
TDKS equations only yields the TDKS non-interacting wavefunction.  The true
wavefunction may differ so dramatically from the KS wavefunction
that observables evaluated on the latter may be inaccurate.

\end{itemize}
The remainder of this essay is a discussion of the various
areas of TDDFT applications and development, and the challenges
presently facing us.  We begin in the middle, with the linear
response regime, where most of the applications presently are,
then go to non-perturbative phenomena, and end with ground-state
applications.

\begin{figure}[htb]
\unitlength1cm
\begin{picture}(6,6)
\put(-2.5,13.5){
\includegraphics{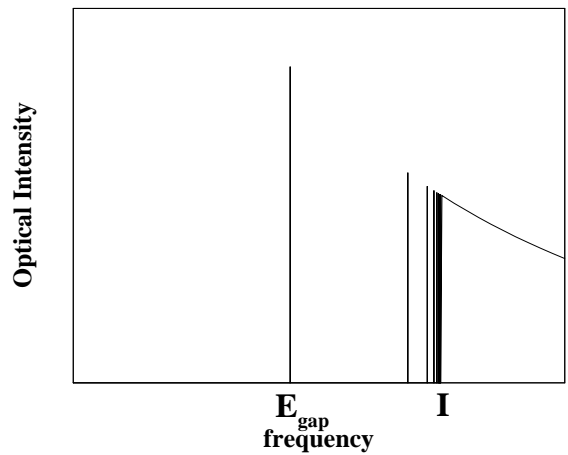}}
\end{picture}
\caption{Same as Fig. \ref{f:exact}, but showing higher frequencies, including
the
infinite Rydberg series of states as the ionization threshold is approached.}
\label{f:exact2}
\end{figure}
We start with applications to the {\bf  excitations of atoms and molecules}.
An important point we wish to emphasize here is the wealth
of prediction made by any TDDFT approximation.
The simplest real system of interacting electrons is the He atom,
and even it has a rich and complex optical absorption spectrum.
Returning to Fig. \ref{f:exact}, we note that a calculation
of the optical spectrum of the bare ground-state KS system,
often\cite{AGB03} looks quite similar to the exact one,
with TDDFT corrections merely shifting and resizing peaks.
In Fig. \ref{f:exact2}, we zoom out a little, and see the
ionization threshold at $\omega=I$
and the infinite Rydberg series of excitations just to its left.
If one calculates the optical response of $N$ non-interacting
electrons in the 
exact KS ground-state potential, i.e., what we call the KS response, 
its ionization threshold is in {\em exactly} the right place,
by virtue of the DFT version of Koopman's theorem\cite{PPLB82}.

From the very earliest calculations of transition frequencies\cite{PGG96,C96},
it was recognized that the inaccuracy of
standard density functional approximations (LDA, GGA, hybrids)
for the ground-state XC potential
leads to inaccurate KS eigenvalues. 
Because the approximate KS potentials have {\bf incorrect asymptotic behavior}
(they
decay exponentially, instead of as $-1/r$), the KS orbital
eigenvalues are insufficiently negative, the ionization
threshold is far too low, and Rydberg states are often
unbound.  This is therefore a {\em ground-state} sin.

Given this disastrous behavior, many 
methods have been developed to asymptotically correct potentials\cite{GGGB02,WAY03}.
Any corrections to the ground-state potential are
dissatisfying, however, as the resulting potential is {\em not} a functional
derivative of an energy functional.  Even mixing one approximation
for $v\xc(\br)$ and another for $f\xc$ has become popular.
A more satisfying route is to use the optimized effective potential (OEP) 
method\cite{GKKG98,UGG95}
and include exact exchange or other self-interaction-free functionals.
This produces a far more accurate
KS potential, with the correct asymptotic behavior. 
The chief error is simply the correlation contribution to the
position of the HOMO, i.e, a small shift.  All the main features
below and just above $I$ are retained.

Why has the poor quality of ground-state potentials
not impeded the rapid growth of TDDFT calculations for excitations
in quantum chemistry?  For many molecules,
the lowest excitations are not Rydberg in character, and the orbitals
do not depend on the large-$r$ behavior of the potential.  
But there are important cases where the problem does show up.  The 
`fruitfly' of TDDFT benchmarks is the $\pi\to\pi^*$ transition
in benzene.  This occurs at about 5 eV in a ground-state LDA calculation,
and ALDA shifts it correctly to about 7 eV\cite{VOC02}.
Unfortunately, this is
in the LDA continuum, which starts at about 6.5 eV!  So how is it
possible to get this right in ALDA?

The answer is that ALDA usually yields good oscillator strengths,
even for states pushed into the continuum\cite{WMB03}.  
The reason is simple, and was suggested long ago in early photoabsorption
calculations by Zangwill and Soven\cite{ZS80}.  The KS LDA potential looks
very much like the exact one (especially in the interior, as the occupied
orbitals yield a good approximation to the true density), shifted
up by a constant, due to the lack of derivative discontinuity\cite{PPLB82}.
The shift pushes the Rydberg states into the continuum, but
retains their contribution to the optical spectrum.  Likewise for the
benzene transition.  Hence ALDA can still be used and trusted for
that transition.  Moral:  {\em Just because it's in the continuum,
doesn't mean it's not right}.

The cartoon of Fig. \ref{f:exact} changes when bonds are stretched.
Of particular interest in biochemistry are {\bf charge-transfer
excitations}, especially between weakly bonded molecules.  Capturing these seems
unnatural within TDDFT, for the simple reason that the numerator
in $\chi$ in Eq. (\ref{chis}) vanishes as the molecular separation
goes to infinity.   Thus, in the density-density response, the 
oscillator strength for these transitions is exponentially small.
Recently\cite{GBb04}, it has been shown how to build an empirical
approximation to an adiabatic $f\xc$ that can capture these
effects, but it is one that {\em grows} exponentially with $|\br-\br'|$.
Thus this is a sin of {\em locality}, in which 
local approximations to $f\xc$ miss a qualitative feature.

A presently open question is the extraction of
{\bf double excitations}\cite{TAHR99,TH00}.
In the adiabatic approximation, these are lost from the
linear response equations.  
When a double excitation lies close to a single excitation,
elementary quantum mechanics shows that $f\xc$ must have a
strong frequency dependence
\cite{MZCB04}.  Thus this problem is due to the
adiabatic approximation: a sin of  {\em forgetfulness}.
Recently, 
a post-adiabatic TDDFT methodology has been 
developed\cite{MZCB04,ZB04} for
including a double excitation when it is close to
an optically-active single excitation, and works
well for small dienes\cite{CZMB04}.
It had been hoped that, by
going beyond linear response, non-trivial double excitations would be
naturally included in, e.g.,
TDLDA, but it has recently been proven
that, in the higher-order response in TDLDA, the double excitations
occur simply at the sum of single-excitations.  Thus we do not currently
know how best to approximate these excitations.  This problem is particularly
severe for quantum dots, where the external potential is parabolic,
leading to multiple near degeneracies between levels of excitation.

\begin{figure}[htb]
\unitlength1cm
\begin{picture}(6,6)
\put(-2.5,13.5){
\includegraphics{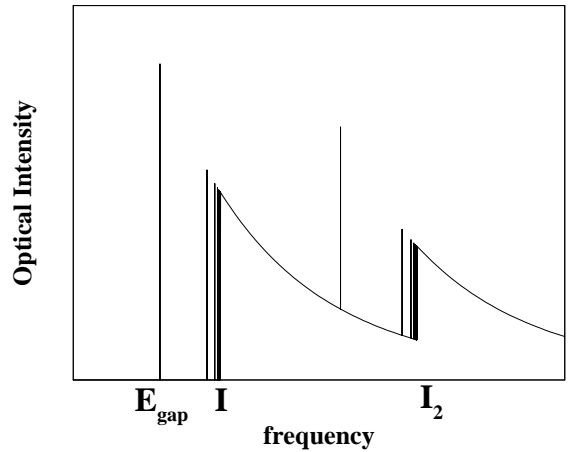}}
\end{picture}
\caption{Same as Fig. \ref{f:exact2}, but now including the second ionization
threshold}.
\label{f:exact3}
\end{figure}
Concerns about both ionization potentials and double excitations
are combined when we consider more of the optical response of the He
atom.
Zooming out just a little more in Fig. \ref{f:exact3}, we
see that there is of course a {\bf second} ionization potential,
when a second electron is stripped off the atom or molecule.
For reference, in a He atom, $I=24.2$ eV, and $I_2=54$ eV.
But the bare KS response contains {\em only} the first threshold.
It has no structure at all in the region of the second ionization.
Our simple density functional approximations to $f\xc$ tend to shift
and resize peak positions.
It is very difficult to imagine {\em density} or orbital functional
approximations to $f\xc$ that can build in the $\omega$-dependence
needed to create a second threshold.

Lastly in this section, we mention recent progress in developing a theory for 
{\bf electron scattering} from molecules.  This was one of the original
motivations for developing TDDFT.   One approach would be to 
evolve a wavepacket using the TDKS equations, but a more direct
approach has been developed\cite{WMB04},
in terms of the response function $\chi$
of the $N+1$ electron system (assuming it is bound).   This uses similar
technology to the discrete transition case.  Initial results for
the simplest case, electron scattering from He$^+$, suggest
a level of accuracy comparable to bound-bound transitions, at least
for low energies (the most difficult case for traditional methods,
due to bound-free correlation\cite{N00}).

A key question that often arises is the need for {\bf time-dependent current
DFT (TDCDFT)}, or not.  The Runge-Gross theorem proceeds by first proving a
one-to-one correspondence between currents and scalar potentials.
Obviously, the current is needed in the presence of time-dependent
magnetic fields, but in their absence, is it necessary?  By
continuity, $dn/dt = -\nabla\cdot \bj(\bx)$, so that the density
is uniquely determined by the current (up to its initial value),
but not vice versa.  It would
seem preferable to stay within the simpler density functional
theory where possible.
A careful examination of the conditions of applicability of the Runge-Gross
theorem to finite systems\cite{GK90} shows that all atoms
and molecules satisfy the necessary conditions of potentials
vanishing sufficiently rapidly as $r\to \infty$.
However, early work showed that the gradient expansion (the origin of
GGA's for the ground state) fails within TDDFT, but behaves
reasonably within the current theory, yielding the Vignale-Kohn
approximation\cite{VK96,VUC97} for the response kernel, which has
frequency dependence.

These questions become relevant to the {\bf optical response of 
bulk insulators}.   The Dyson-like Eq. (\ref{Dyson}) becomes a
matrix equation with indices of the reciprocal lattice vectors ${\bf G}$
for each perturbation of wavevector $\bq$.  As $q\to 0$, to find the
optical response, any local approximation to $f\xc$ produces a negligible
correction to the RPA response ($f\xc=0$), as the Hartree contribution
(correctly) blows up as $1/q^2$.  Thus, to have a noticable effect, 
the XC kernel
must have a $1/q^2$ component as $q\to 0$.  While this effect is sometimes
referred to as 'ultra'-non-local, we prefer to call it simply non-local,
as the range of non-locality is precisely that of the Hartree contribution.
The optical response of the solid can be found within TDDFT by perturbing
the system with a long-wavelength perturbation of wavevector $q$, and by
carefully taking $q\to 0$.  This requires
extending the RG theorem to periodic Hamiltonians\cite{MSB03}.
On the other hand, a $q=0$ calculation, with just
the period of the lattice, is possible within TDCDFT, and the non-local
contribution in TDDFT appears as a local contribution
within TDCDFT, with no unusual non-locality needed in the
current density.  For example, the VK approximation produces a finite
correction, whereas LDA and GGA do not.   Thus, in this
and other cases, TDCDFT is not strictly necessary, but provides a 
more direct description and a route to extract quantities
that are non-local in TDDFT.
Similar observations apply to the {\bf polarizabilities of long
organic polymers}.  ALDA and GGA greatly overestimate these quantities,
but VK often does much better. 

In this context, an important 
challenge is the proper description of {\bf excitonic peaks} in the optical
spectra of insulators. It was recently demonstrated\cite{ORR02,SOR03,MSR03}
that with complicated orbital-dependent approximations for $f\xc$, which
were derived from the Bethe-Salpeter equation, excitonic effects can be
described perfectly. However, the presently available
schemes require a GW calculation in the first place. There remains
the challenge to find {\it sufficiently simple} (possibly current-dependent)
approximations that are able to describe excitonic effects. 

Another interesting question in the optical response of insulators
is the one of the {\bf gap}.  It is well-known from ground-state DFT that
the gap in the spectrum of KS eigenvalues (the KS gap) differs
from the true gap by a quantity called the derivative discontinuity\cite{DG90}.
Ignoring excitons within the gap, shouldn't TDDFT correct the KS
gap to yield the true gap?  The answer is, yes, but again the XC
kernel that does this must be very sophisticated, just as in
our double ionization example.  Since $\chi\s$
develops an imaginary part for frequencies above the KS gap, the kernel
must have a branch cut that exactly suppresses this in order to
widen the gap.
There is a close analogy to the problem of charge-transfer excitations:
Remove an electron from the 
donor to infinity.
This costs the ionization energy $I_{\mathrm{DON}}$. Then move the electron 
from infinity to the acceptor.
In this way one gains the energy $-A_{\mathrm{ACC}}$. So the excitation energy is
$\Delta E = I_{\mathrm{DON}}
- A_{\mathrm{ACC}} = \epsilon^{\mathrm{LUMO}}_{\mathrm{ACC}}
- \epsilon^{\mathrm{HOMO}}_{\mathrm{DON}} + 
\Delta\xc^{(\mathrm{ACC})}$, where $\Delta\xc^{(\mathrm{ACC})}$ is the 
discontinuity in the ground-state $v\xc(\br)$ of the acceptor.
This formula is reminiscent of the band gap in insulators\cite{DG90}. 
Furthermore, the next-order correction is the
Coulomb-interaction between the electron on the 
acceptor and the hole on the donor which,
in solids, corresponds to the exciton binding energy. 

On the boundary between extended systems and molecules is
{\bf transport through single molecules} connected to bulk metal
leads\cite{NR03}.  There is enormous interest in this as a key component
in future nanotechnology.  Present formulations use ground-state
density functionals to describe the stationary non-equilibrium
current-carrying state\cite{BMOT02}.  But several recent suggestions consider
this as a time-dependent problem\cite{SA04,VT04,GCb04,KSARG05,BCG05},
and use TD(C)DFT for a full description
of the situation.  Only time will tell if TDDFT is really needed for 
an accurate description of these devices.
In the special case of weak bias, XC corrections to the Landauer
formula are missed by local approximations, the sin of {\em locality}
\cite{BE05}.

Next we turn our attention to beyond-perturbative regimes.  Due to advances
in laser technology over the past decade, many experiments are
now possible in regimes where the laser field is stronger than the
nuclear attraction.  There are a whole host of phenomena that TDDFT
might be able to predict:  high harmonic generation, multi-photon
ionization, above-threshold ionization, above-threshold dissociation, etc.  
For {\bf high harmonic generation}, TDDFT calculations
have been rather succesful for atoms \cite{UEG96,EG99} and
 molecules \cite{CC01,BNZM03}.
In the near future, this might become very important for the generation 
of atto-second laser pulses \cite{CBY94,CMK97,PS99}. 
For {\bf multi-photon ionization}, the
relative proportion of double to single ionization for He, while given
much better in approximate TDDFT calculations than in previous
calculations assuming a sequential mechanism, still does not
show the same pronounced features (the 'knee') seen in experiments \cite{PG99,BC01}.
The electron spectra from {\bf above-threshold ionization}
have recently been calculated within TDDFT \cite{PRS00,NBU04}.
Since the ionization yields and above-threshold ionization spectra depend on probabilities extracted directly
from the wavefunction, these errors are suspected to be sins of
the {\em wavefunction}, mentioned above.
An important task for the future
will be the design of more realistic expressions for ionization
probabilities or, more generally, transition probabilities as functionals
 of the time-dependent density or the time-dependent KS orbitals. 
First steps in this direction can be found in Ref. \cite{PG99}.

  While the need for more accurate KS
potentials was first noticeable in calculating
excitations, it is even more acute in strong laser fields.  To
ensure an atom or molecule ionizes accurately in an approximate
TDKS calculation, Koopmans' theorem ($I=-\epsilon_{HOMO}$)
should be well-satisfied,
and this again requires using OEP exact
exchange\cite{GKKG98,UGG95} or other self-interaction-free functionals.

The field of {\bf quantum control} has, so far, mainly concentrated on 
manipulating the motion of the nuclear wave packet on a given set of
precalculated potential energy surfaces, the ultimate goal being the 
femto-second control of chemical reactions \cite{RZ00}. 
With atto-second pulses available, the control of electronic dynamics has come within reach. A marriage
of optimal-control theory with TDDFT appears to be the ideal 
theoretical tool to tackle this situation. However, it will 
bring with it its own difficulties
and challenges for approximate functionals.  Using the functional
algorithms developed by Rabitz and others \cite{ZBR98,TKO92}, we
can find the optimal
pulse that drives a He atom from its ground state to its first excited
state, $1s2p$.   (In practice, we do not reach exactly 100\% occupation, due
to a finite penalty factor).   

Now repeat this experiment
on non-interacting electrons sitting in the same potential.  
Such a pulse cannot be found, i.e., the non-interacting system
is not controllable, whereas the interacting system is. 
The two non-interacting electrons must follow the
same time evolution as they start from the same initial 1s orbital and are
exposed to the same laser field. Hence the time-dependent wave function of 
the two non-interacting electrons must have the form:
\begin{equation}
\Phi(r_1,\sigma_1,r_2,\sigma_2,t) =
    \varphi(\br_1,t)\varphi(\br_2,t)\chi_S (\sigma_1,\sigma_2)
\end{equation}
where $\chi_S (\sigma_1,\sigma_2)$ represents the (antisymmetric)
spin-singlet part of the wavefunction.
But we want to maximize the occupation
\begin{equation}
\left| < \Phi(T) | \Phi_{1s,2p} > \right|^2
\end{equation}
of the time-propagated wavefunction $\Phi(T)$ at the end, $T$, of
the laser pulse in the lowest excited state
\ben
\Phi_{1s,2p} = \frac{1}{\sqrt{2}}
(\varphi_{1s} (\br_1) \varphi_{2p} (\br_2) +
\varphi_{2p} (\br_1) \varphi_{1s} (\br_2) ) \chi_S (\sigma_1,\sigma_2)
\end{equation}
Expanding the final wavefuntion $\Phi(T)$ in the complete set  of
single-particle orbitals representing the eigenfunctions of the
unperturbed system, one sees that the best possible
occupation is $50\%$.

If TDDFT is used to describe the multi-electron dynamics, how would one 
properly define the control target, i.e., the functional to be maximized? Choosing,
as control target, the overlap with an excited-state Kohn-Sham determinant does not
seem to be a good idea in view of the above dilemma. If, on the other hand, the
time-dependent density of a fully controlled $(1s)^2$ to $(1s,2p)$ transition of
the interacting system is given, would an exact TDKS calculation reproduce such an
optimal evolution?  The answer is yes,  but $v\xc(\br t)$
must be very special to do so.  To see this, take the density evolution
from the exact Schr\"odinger equation, and invert the TDKS equation for the
single (doubly-occupied)
time-dependent orbital.  The final state KS potential is very odd,
producing the density of two orbitals of different symmetry from
a single doubly-occupied orbital\cite{MWB01}.

For a subset of cases in which molecules are exposed to strong fields,
the nuclear motion can be treated classically. The electrons then feel 
the Coulomb field of classically moving nuclei as well as the laser field.
In this case, the electronic motion is well described by ordinary TDDFT. 
However, {\bf when nuclear trajectories split}, e.g., when a molecule
has a 50\% chance of dissociation in a given laser pulse,
the classical treatment fails.
A multicomponent TDDFT\cite{KG01,KGLG03,KLG04}
has been developed for
electrons and nuclei which, in principle, handles such a situation.
In practice, one needs to develop appropriate approximations that can
build in the appropriate physics of, e.g., dissociating nuclei.  
Again, in this case, orbital-dependent
functionals appear crucial\cite{KLG04,KG01}.

Finally, and fondly, we return to the ground state.
The last general application mentioned was the odd-sounding process
of {\bf using TDDFT to generate ground-state approximations},
via Eq. (\ref{Exc}).  By inserting an approximation for $f\xc$, we
get an approximation to $E\xc$.
Most importantly for biological systems, Eq. (\ref{Exc})
provides a route to {\bf  van der Waals} forces
for separated pieces of matter, and so is being much studied by developers.
In particular, the coefficient in the decay of the energy between
two such pieces ($C_6$ in $E\to -C_6/R^6$, where $R$ is their separation)
can be accurately (within
about 20\%) evaluated using a local approximation to the frequency-dependent
polarizability\cite{ALL96,ALLb96,OGSB97,KMM98}.
Recent work shows that the response functions of TDDFT
can yield extremely accurate dispersion energies of monomers\cite{MJS03}.
More recently, the frequency integral in Eq. (\ref{Exc})
has been done approximately, yielding an
explicit non-local density functional\cite{DRSL04} applicable at all
separations.

One can go the other way, and try using Eq. (\ref{Exc}) for all
bond lengths\cite{F01,FG02}.
Such calculations are costly, as the functional is very high up on
Jacob's ladder of sophistication, including both occupied and unoccupied
KS orbitals\cite{PRTS04}.  However, they have
the merit of being entirely non-empirical
and, where successful, can be used as a starting point for new approximations.
In fact,  Eq (\ref{Exc}) provides a KS 
density functional that allows {\bf bond-breaking} without artifical symmetry
breaking\cite{FNGB05}.
In the paradigm case of the H$_2$ molecule, the binding energy
curve has no Coulson-Fischer point, and the dissociation occurs correctly
to two isolated H atoms.  Unfortunately, simple approximations, while
yielding correct results near equilibrium and at infinity, produce an
unphysical repulsion at large but finite separations.
This can be traced back\cite{FNGB05} to the
lack of double excitations in any adiabatic $f\xc$.

We end with a {\bf summary}.  Most importantly, TDDFT has become
extremely popular as a method for calculating electronic
excited-state energies in chemistry.  In this arena, it has
become as robust (or as flaky, depending on your perspective) as
ground-state DFT, and is being used to really understand the
photochemistry of many interesting problems.  Furthermore,
our favorite crude density functional
approximations from the ground-state serve well here.
We are very thankful for this, and it has led to tremendous interest
in further methodological development.

In principle, however, TDDFT yields predictions for an enormous
variety of phenomena, and electronic excitations are only the tip
of an iceberg.  We have mentioned a few.  Even limiting
ourselves to linear response, there are double excitations, second ionization
thresholds, optical
response of solids, gaps in solids, transport through single molecules.
Combining with the fluctuation-dissipation theorem, TDDFT yields
a route to van der Waals forces and bond-breaking with symmetry problems.
In strong fields, there are high harmonic generation, multi-photon ionization,
above-threshold ionization, quantum control, and quantum nuclear motion.

For some of these areas, simple application of density functionals within
the adiabatic approximation, works well, but for many, such methods miss
some qualitative features (e.g., double excitations, or non-locality
in the response of solids).  A now standard
step upward in sophistication is to use orbital-dependent functionals
(at least, among developers),
and these
cure some of the difficulties (e.g., the first ionization threshold or the
polarizability of long-chain polymers).  But such functionals are unlikely
to cure all of the problems (e.g., inclusion of double excitations or
defining the target in quantum control) for
properties that are of interest experimentally and technologically.
We happily look forward to many interesting years of development to come.

We thank Maxime Dion, Vazgen Shekoyan,
and Adam Wasserman for useful discussions.
K.B. gratefully acknowledges support of the US Department of Energy, under
grant number DE-FG02-01ER45928.
This work was supported, in part, 
by the Deutsche Forschungsgemeinschaft, the EXC!TiNG Research and Training
Network of the European Union and the NANOQUANTA Network of Excellence. 
Some of this work
was performed at the Centre for Research in Adaptive Nanosystems
(CRANN) supported by Science Foundation Ireland (Award 5AA/G20041).

\end{document}

%% file: dft.tex
\def\bea{\begin{eqnarray}}
\def\eea{\end{eqnarray}}
\def\ben{\begin{equation}}
\def\een{\end{equation}}
\def\benu{\begin{enumerate}}
\def\enu{\end{enumerate}}

\def\n{n}

\def\sss{\scriptscriptstyle\rm}

\def\g{_\gamma}

\def\l{^\lambda}

\def\1var{(\bx_1...\bx\N)}

\def\half{\frac{1}{2}}

\def\br{{\bf r}}

\def\bx{{\br t}}

\def\bq{{\bf q}}
\def\bj{{\bf j}}

\def\s{_{\sss S}}
\def\xc{_{\sss XC}}

\def\Hxc{_{\sss HXC}}

\def\N{_{\sss N}}

\def\ext{_{\rm ext}}

\def\adia{^{\rm adia}}

\def\sph_int{ {\int d^3 r}}
